\documentclass[apjl]{emulateapj}
\usepackage{apjfonts}
\usepackage{graphicx}
\usepackage{amsmath}
\usepackage{amssymb}
\usepackage{amsfonts}
\shorttitle{Hard Emission from Cen A}
\shortauthors{Burke et al.}

\begin{document}
\newcommand{\lerg}[1]{10^{#1}~{\rm erg~s^{-1}} }
\newcommand{\msol}{{\rm ~M_\odot}}
\newcommand{\mdot}{\dot{M}}
\newcommand{\nh}[1]{10^{#1}~{\rm cm^{-2}} }

\title{The hard X-ray continuum of Cen A observed with INTEGRAL SPI}

\author{Mark~J.~Burke\altaffilmark{1},
Elisabeth Jourdain\altaffilmark{1},
Jean-Pierre Roques\altaffilmark{1}, and 
Daniel A. Evans\altaffilmark{2}
}

\altaffiltext{1}{CNRS, IRAP, 9 Av. colonel Roche, BP 44346, F-31028 Toulouse cedex 4, France
\email{mburke@irap.omp.eu}}

\altaffiltext{2}{Harvard-Smithsonian Center for Astrophysics, 60 Garden Street, Cambridge, MA 02138}

\begin{abstract}
We revisit the average hard X-ray spectrum from the AGN of Centaurus A (Cen A) using ten years worth of observations with \emph{INTEGRAL} SPI.  This source has the highest flux observed from any AGN in the SPI bandpass (23 keV--8 MeV). The 10 year lightcurve of Cen A is presented, and hardness ratios confirm that the spectral shape changes very little despite the luminosity varying by a factor of a few.   

Primarily, we establish the presence of a reflection component in the average spectrum by demonstrating an excess between $20-60$ keV, from extending the spectral shape observed at low energy to the SPI regime.  The excess in \emph{Chandra} HETGS and \emph{INTEGRAL} SPI data is well described by reflection of the dominant power law spectrum from a neutral, optically-thick atmosphere.  We find that the reprocessed emission contributes $20-25\%$ of the $23-100$ keV flux.

The existence of a cut-off at tens to hundreds of keV remains controversial. Using simulated spectra, we demonstrate that a high energy cut off reproduces the observed spectral properties of Cen A more readily than a simple power law. However, we also show that such a cut-off is probably underestimated when neglecting (even modest) reflection, and for Cen A would be at energies $>700$ keV, with a confidence of $>95\%$.   This is atypically high for thermal Comptonizing plasmas observed in AGN, and we propose that we are in fact modelling the more gradual change in spectral shape expected of synchrotron self-Compton spectra.


\end{abstract}


\section{Introduction}
\label{s:intro}

The radio galaxy Centaurus A (Cen A, NGC 5128), at a distance of 3.8 Mpc \citep{2010PASA...27..457H} is our nearest large early-type galaxy and one of the most studied active galactic nuclei (AGN).  Despite this, we are still able to say comparatively little about the spectral properties of the AGN in the hard X-ray to $\gamma$-ray regime.  Outstanding questions relate to the presence of reprocessed emission in the 20-100 keV spectrum and the existence of a high energy downturn in the spectral shape at a few hundred keV.  

In the soft X-ray, \emph{Chandra} and \emph{XMM Newton} grating observations have been used to demonstrate that a moderately absorbed ($N_H \approx \nh{23}$) power law component dominates the $0.1 -10.0$ keV spectrum. However, inspection of the fit residuals indicated the need for extra components, the most prominent of which was a narrow ($EW \approx 10-50$ eV) emission line consistent with neutral Fe $\rm{K\alpha}$ \citep{2004ApJ...612..786E}.  It is feasible that the excess emission above the power law arises from reflection of the central X-ray source from surrounding material \citep{1991MNRAS.249..352G}.  Indeed, the absence of more highly-ionised iron (i.e. Fe XXV, Fe XXVI) and the narrowness of the  Fe  $\rm{K\alpha}$ line profile is indicative that reflection is occurring far from the innermost portions of the accretion disk \citep{2004ApJ...612..786E}. The presence of reflection in the spectrum should lead to an apparent excess $\gtrsim 20$ keV \citep[the so-called `Compton hump',][]{1991MNRAS.249..352G}. 

The existence of a high energy cut-off at $E_{cutoff}$ would have profound implications for the accretion mechanism of the Cen A AGN, and is the subject of controversy.  Fundamentally it would inform on the origin of the primary emission component; whether it is produced in the jet \citep{1992ApJ...397L...5M,1996ApJ...461..657B} by synchrotron self-Compton (SSC) \citep{1969ARA&A...7..375G}  or near to the accretion disc from thermal comptonisation \citep{1993ApJ...416..458D,1994ApJ...421..153S}.  For the case of thermal comptonisation, a high energy cut-off $E_{cutoff}$ in the spectrum will exist at $2-3$ times the temperature of the plasma $kT_e$ \citep{2001MNRAS.328..501P,2009MNRAS.399.1293M}, whereas for SSC the narrow-band spectral shape changes more gradually between the keV and GeV regimes. \cite{2013MNRAS.433.1687M} show that a reasonable upper-limit to the average $kT_e$ for all AGN corresponds to $E_{cutoff}\approx 200$ keV, based on cosmic diffuse background measurements \citep{2007A&A...463...79G}.  The SSC scenario is not new in the context of Cen A \citep{2001MNRAS.324L..33C}, and describes the spectral energy distribution (SED) well at soft X-rays (observed by \emph{SWIFT}) and the GeV $\gamma-rays$ \citep{2010ApJ...719.1433A}.  Cen A is an FR I radio galaxy \citep{1974MNRAS.167P..31F}, which are believed to be misaligned BL Lac objects, i.e. the jet is not aligned with the line-of-sight.  However, FR I galaxies are more luminous in $\gamma$-rays than expected for blazars viewed off-axis, suggesting emission from a slower moving region of the jet (either at a greater distance from the nucleus~\citep{2003ApJ...594L..27G}, or in a so-called `sheath' around a faster-moving inner-jet \citep{2000A&A...358..104C}), as the beaming angle is inversely proportional to the bulk Lorentz factor.  The SED of some other FR I galaxies are also well-described by SSC emission \citep{2009ApJ...699...31A,2009ApJ...707...55A}.

The Compton hump from reflection was first detected in Seyfert galaxies using the \emph{Ginga} satellite \citep{1994MNRAS.268..405N}, but its discovery in the Cen A spectrum remains tentative.   \cite{2007ApJ...665..209M} used \emph{Suzaku} observations to claim a non-detection, while \cite{2011ApJ...743..124F} demonstrated a residual excess in the continuum  between $10-20$ keV but were not able to compensate for this in their reflection modelling (their figure 5), neither do they adequately model the emission \emph{above} 20 keV.  \cite{2011A&A...531A..70B} used \emph{INTEGRAL} IBIS and SPI data to claim a $1.9\sigma$ detection of the reflection component. However, their approach  merely assessed the consistency of the flux from a best-fit reflection component with zero, which is not a valid method to test for the presence of a spectral component \citep[see the discussion in][]{2002ApJ...571..545P}.  Moreover, the Compton hump must be consistent with the inferred reflection properties observed in the soft X-ray bandpass and with the input spectrum being from the central source.

Observations of Cen A with \emph{Fermi-LAT} have shown that the GeV spectrum follows a much steeper power law \citep[$\Gamma\sim2.5$][]{2010ApJ...719.1433A} than is measured in the X-ray regime \citep[$\Gamma\sim 1.6-1.9$,][]{2004ApJ...612..786E,2011A&A...531A..70B,2011ApJ...733...23R,2011ApJ...743..124F}.  This difference in spectral shape between the keV and GeV regime might indicate the presence of a high energy cut-off, as would be expected for the thermal Compton scenario, or it could be the case that the spectrum evolves gradually between low and high energy, as is expected for SSC spectra.   Previous work by \cite{1995ApJ...449..105K} used OSSE data to claim a high energy cut-off with $E_{cutoff}$ inversely proportional to intensity over the 120-700 keV band.    \cite{2011ApJ...733...23R} used the highest quality datasets from 12 years of \emph{RXTE} HEXTE data to constrain $E_{cutoff}>2$ MeV during the brightest epochs, while \cite{2011A&A...531A..70B} used \emph{INTEGRAL} JEM-X/IBIS/SPI/PICsIT data to fit an absorbed cut-off power law with a best-fit $E_{cutoff}=434^{+106}_{-73}$ keV, and subsequently re-fit the spectrum with a thermal comptonisation model (\emph{compPS}), finding $kT_e=206\pm62$ keV.

The present paper makes use of 10 years worth of data obtained with the primary spectrometer, SPI \citep{2003A&A...411L..63V,2003A&A...411L..91R}, on-board the {\bf Int}ernational {\bf G}amma {\bf R}ay {\bf A}strophysics {\bf L}aboratory \citep[\emph{INTEGRAL},][]{2003A&A...411L...1W}.  At present SPI is the optimal instrument for studying both AGN reflection and cut-off components simultaneously, as it is sensitive over the 20 keV -- 8 MeV range with an excellent spectral resolution (2.5 keV at 1.3 MeV).  In this paper we study the average Cen A spectra, and present evidence of the reflection component in the SPI data based on extending the best-fit continuum of \emph{Chandra} grating observations to higher energies.  Subsequently, we use simulated spectra to carry out a robust statistical test for the existence of a high-energy cut-off.  Finally, we investigate the extent to which we can accurately recover $E_{cutoff}$ with simple models when reflection is also present in the spectra of AGN.







\section{Preparation and Analysis}

\subsection{Data Reduction}
\label{sec:datared}
We select only SPI observations where the Cen A nucleus is inside the $\pm 10^\circ$ FoV. These data are further reduced by excluding science windows with a high background activity due to passage of the observatory through the Earth's radiation belts or increased solar activity. The source flux extraction for each energy band is performed through a model fitting procedure which compares the source and background fluxes convolved with the instrument transfer matrix with the count rates recorded in the detector plane \citep[as outlined by][]{2009ApJ...704...17J}. We model the sky based on a point source at the location of Cen A, and assume the flux to be constant over one revolution ($\sim 3$ days, or less if multiple targets were observed during that revolution) of the spacecraft.  Each revolution is comprised of numerous `science windows', lasting $2-4$ ks, which commence as the observatory is re-pointed as part of the dithering strategy.  We model the background as an empty field that is variable  on a timescale of 5 science windows for a given revolution.  This reduction was carried out using the SPI Data Analysis Interface (\emph{SPIDAI}\footnote{http://sigma-2.cesr.fr/integral/spidai}).

After data reduction the average SPI spectrum of Cen A has an effective exposure time of 1.355 Ms.

Observations between $0.5-10$ keV using the \emph{Chandra} ACIS instrument have shown that the inner $5\arcmin$ of Cen A boasts a wealth of X-ray features besides the bright nucleus, such as the extended jet \citep{2008ApJ...673L.135W}, hot gas \citep{2008ApJ...677L..97K}, shock-lobes \citep{2009MNRAS.395.1999C} and a large population of X-ray binaries \citep{2009ApJ...701..471V,2013ApJ...766...88B}.  These other features are potentially problematic for studying the high-energy emission of the nucleus with SPI, which has an angular resolution of $2.5^\circ$.  However, \emph{Suzaku} observations of Cen A have shown that the  contribution to the $2-10$ keV flux of Cen A from X-ray binaries is $<8\times \lerg{39}$ \citep{2011ApJ...743..124F}, less than $1\%$ of the AGN emission, while the bright extended ($\sim 3\arcmin$) jet \citep{2008ApJ...673L.135W} only contributes $\approx \lerg{39}$.  As we expect the contribution of other X-ray features to diminish at higher energy, we are confident that the emission detected by SPI comes overwhelmingly from the nucleus.  In this paper we make use of reduced and processed \emph{Chandra} spectra presented by \cite{2004ApJ...612..786E}.  

\begin{figure*}
\begin{center}
\includegraphics[width=0.8\hsize]{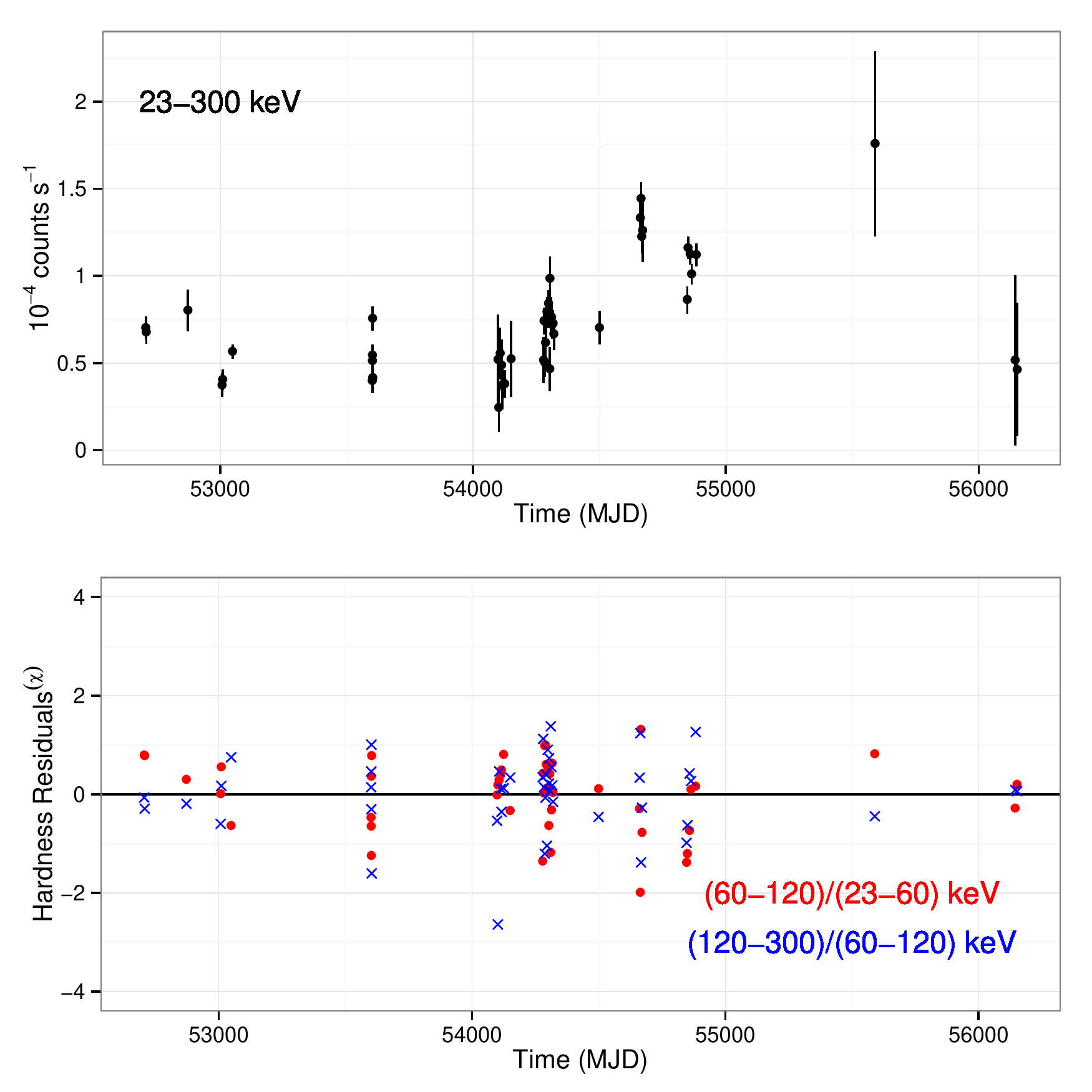} 
\caption{Lightcurve of Cen A observed by \emph{INTEGRAL} SPI (upper panel) and hardness residuals after subtraction of the weighted mean (lower panel).  \label{fig:lc}}
\end{center}
\end{figure*}

\subsection{Analysis} \label{sec:analysis}
\subsubsection{Intensity and Hardness}
\cite{2011ApJ...733...23R} established that the shape of the high-energy spectrum of Cen A does not change with time over the RXTE bandpass, despite the flux varying by a factor of a few.  We tested this over the wider bandpass of the SPI data, by calculating hardness ratios and judging their consistency with a constant value.  The $23-300$ keV lightcurve of the Cen A nucleus is presented in the top panel of figure~\ref{fig:lc}, beneath which we show the residuals of two hardness ratios subtracted from their weighted means.  These support the conclusion of \cite{2011ApJ...733...23R}, i.e. that in the hard X-rays the spectral shape is unchanged despite significant variation in flux.  This consistency of spectral shape allows us to examine the average spectrum of Cen A over all observations.

\subsubsection{Spectral Fitting}
\label{sec:specfit}

Using XSPEC 12.7 we find that the average \emph{INTEGRAL} spectrum of Cen A fitted with an absorbed powerlaw with absorption fixed at $N_H=\nh{23}$  is adequately described ($\chi^2_\nu=46/48$) by a power law of photon index $\Gamma=1.81\pm0.03$.  All spectral fitting uncertainties in this paper are quoted at the $90\%$ level. 

As a further test of spectral consistency, and to demonstrate the validity of using the average Cen A spectrum, we extracted spectra from high and low intensity regimes, defined as being above or below $10^{-4} {\rm counts~s^{-1}}$ (see figure~\ref{fig:lc}). We found nearly identical power law fits, barring a change in normalisation, with $\Gamma_{high} = 1.83$, $\Gamma_{low}=1.81$; within the uncertainty of $\pm0.07$ associated with each measurement.   
 
\begin{figure*}
\begin{center}
\includegraphics[width=0.49\hsize]{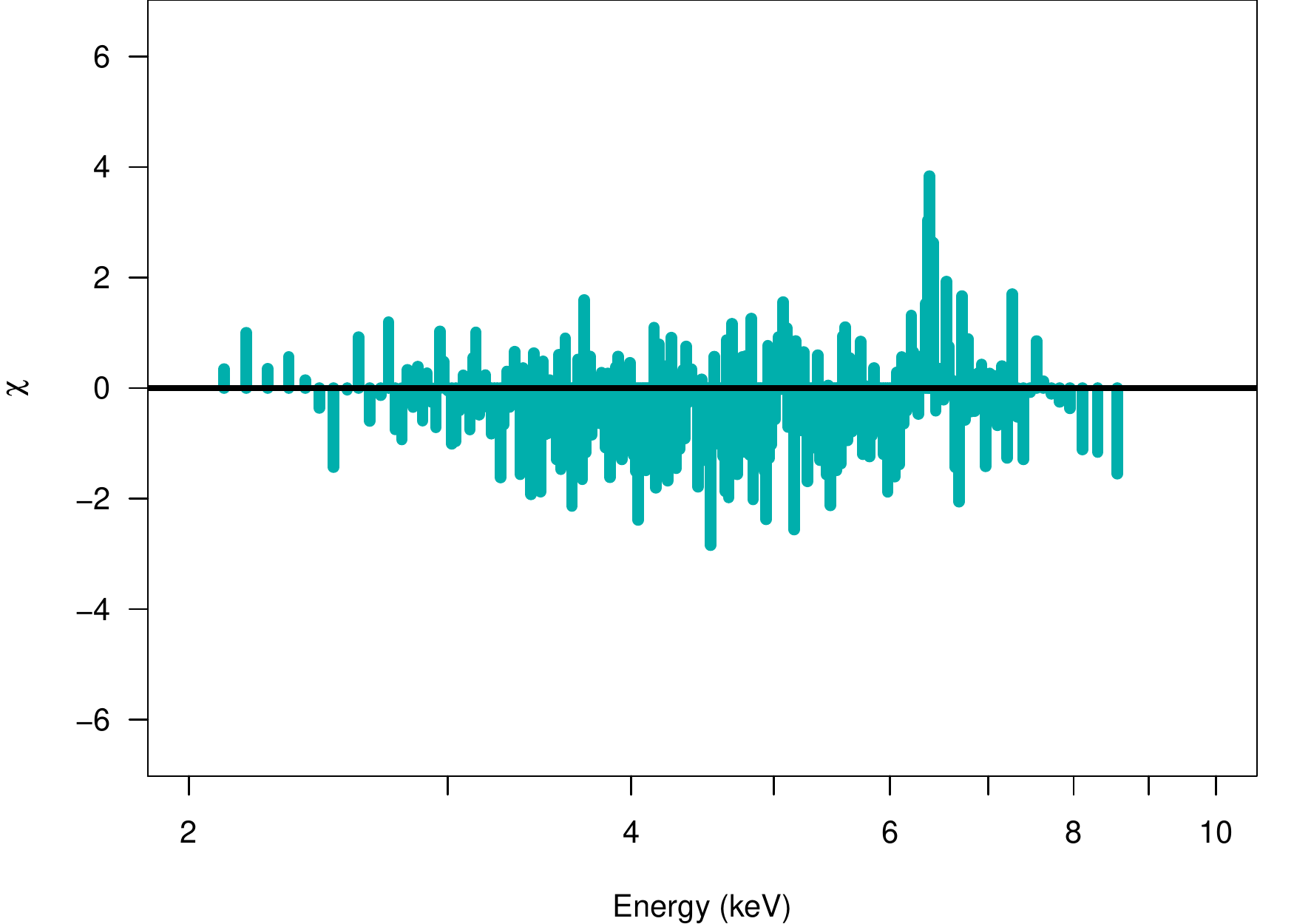} \includegraphics[width=0.49\hsize]{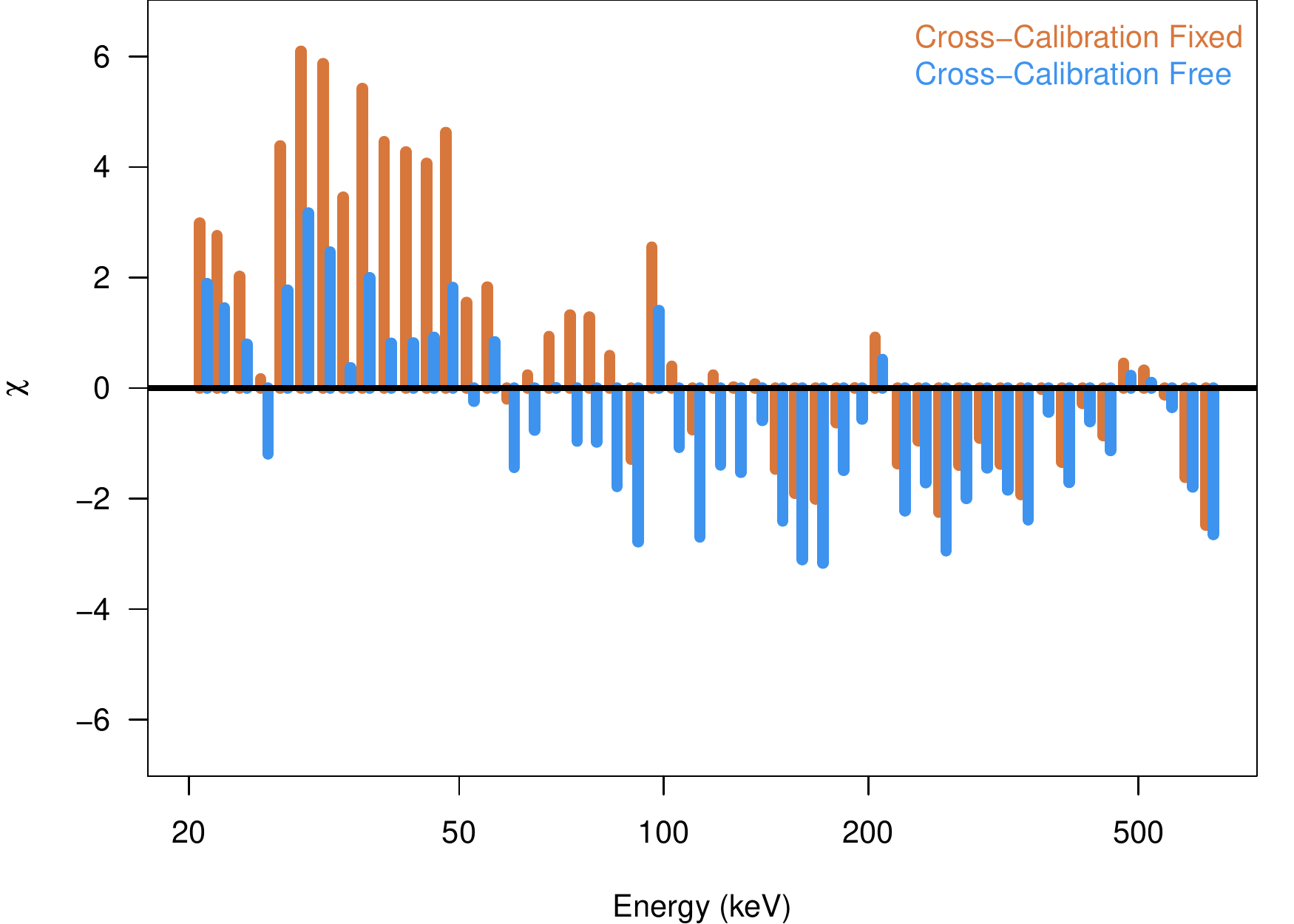}
\caption{The residuals from the dominant power law component found by \cite{2004ApJ...612..786E} using the \emph{Chandra} gratings data, with normalisation free.  On the left are these same \emph{Chandra} data, while the right panel shows the residuals of the average spectrum observed by \emph{INTEGRAL} SPI.  The SPI residuals correspond to having the cross-calibration factor fixed ($C=0.78$, the value found from fitting a single absorbed power law to both datasets), and as a free parameter ($C\sim 0.9$).  For clarity, these instances are offset in $x$.\label{fig:powerlawres}}
\end{center}
\end{figure*}

To evaluate the existence of other model components we assess the consistency of the spectral shape between low and high energy, particularly because low energy data is comparatively good quality.  We revisit the \emph{Chandra} METGS and HETGS spectra of the Cen A core (obsID 1600) presented in \cite{2004ApJ...612..786E} with the same spectral grouping ($>100$ counts per bin).  They found the continuum to be dominated by a $\Gamma\sim1.65$ power law.   We fitted an absorbed power law ({\bf phabs $\times$ powerlaw} in XSPEC) to the \emph{Chandra} spectrum with $\Gamma=1.65$, with the normalisation free to vary. We then introduced the SPI spectrum to gauge how well the spectral shape extended to high energies.  To account for calibration and non-simultaneity between the \emph{Chandra} and \emph{INTEGRAL} observations, we allow a multiplicative factor  $C$ to the INTEGRAL portion of the model.  To determine a reasonable value for C, we first fit the \emph{Chandra} data with an absorbed power law, with a Gaussian to account for the narrow emission in the vicinity of the Fe K complex.  Freezing all key parameters except normalisation prior to adding the SPI data, i.e. such that only the normalisation and $C$ may vary, finding $C=0.78$.  Subsequently, we free the photon index to vary as well, recovering $C=0.9$ (with $\Gamma = 1.71\pm 0.02$).  In figure~\ref{fig:powerlawres} we present the $\chi$ residuals after fitting the simple power law model with $C$ fixed at 0.78 and $\Gamma=1.65$ \citep[the main power law component found by][]{2004ApJ...612..786E}.  In the right-hand panel -- the \emph{INTEGRAL} SPI residuals -- we also present residuals for $C=0.9$.  

In the \emph{Chandra} panel of figure~\ref{fig:powerlawres} the known (unmodelled) reflection component can clearly be seen as an excess in the spectrum at $\sim6.4$ keV.  The systematic shape of the residuals in the right-hand panel, from the SPI data, shows that the $\Gamma=1.65$ power law alone is an inadequate representation of the spectrum at high energy, and we suggest that the apparent excess of the data in the 20-50 keV range and excess of the model at higher energies is the result of the unmodelled reflection component, consistent with the excess at 6.4 keV.

The {\bf reflionx} model \citep{2005MNRAS.358..211R} describes the reflection of a power law spectrum from an optically thick atmosphere. The key parameters of {\bf reflionx} are the photon index $\Gamma$ of the ionising continuum (thus should be tied to the $\Gamma$ of the a separate power law component during fitting), ionisation parameter $\xi$ and the abundance of iron $A_{Fe}$ in the reprocessing medium. Neutral reflection in a distant absorber is a nearly ubiquitous feature of AGN spectra, and  \cite{2013MNRAS.428.2901W} successfully modelled this  using {\bf reflionx}  for a sample of high-quality \emph{Suzaku} observations of AGN (with $\xi$ fixed to the minimal value, as required of neutral material).  They found that their reflection treatment had the required flexibility to reproduce all of the broadband spectra consistently over their sample. Therefore we chose to use {\bf reflionx} in the current work as our preferred model of reflection from AGN.

\begin{figure*}
\begin{center}
\includegraphics[width=0.8\hsize]{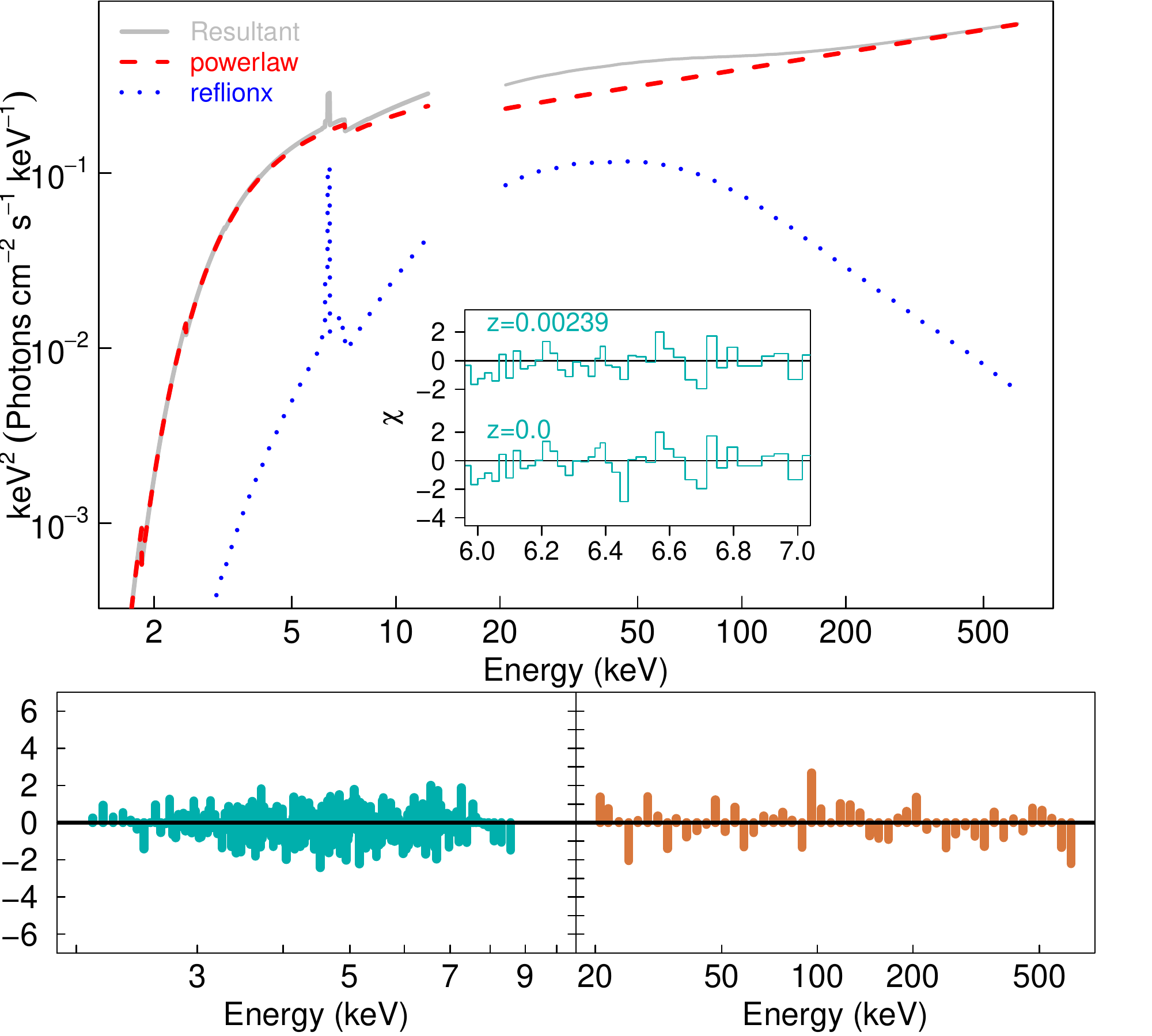} 
\caption{Spectral model including reflection component (top panel), together with the best fit residuals (bottom panels, as in figure~\ref{fig:powerlawres}. Inset we show the effect on the residuals of leaving $z$ fixed to 0 and then having it as a free parameter.\label{fig:spectralfit}}
\end{center}
\end{figure*}

In figure~\ref{fig:spectralfit} we present our best-fit model and residuals for the absorbed power law plus reflection model to the \emph{Chandra} and \emph{INTEGRAL} SPI data (see table~\ref{tab:results}).  During the fitting process we tied the photon index of the reflection component to that of the power law, to be consistent between input and reflected spectrum.  In the first instance, we left $\xi$ free to vary, with $A_{Fe}$ and $z$ frozen at the default values (solar and zero, respectively) and found that it tended towards low values between 10 (the lowest ionisation state) and  an upper-limit of 55, as expected for reflection from a distant surface.  We subsequently followed the treatment of \cite{2013MNRAS.428.2901W} and fixed $\xi =10$.  

We also probed the sensitivity of the fit to $A_{Fe}$ and $z$.  Letting these parameters vary yielded best-fit values of $A_{Fe}=0.69^{+0.24}_{-0.21}$ and $z=2.39^{+1.93}_{-1.41}\times 10^{-3}$.  This value of redshift is fully consistent with that found in the $2M++$ catalogue \citep[$1.826^{+0.017}_{-0.017}\times 10^{-3}$,][]{2011MNRAS.416.2840L}, and henceforth we fixed $z$ to the catalogue value during fitting.  As an inset to figure~\ref{fig:spectralfit}, we include residuals of the $6-7$ keV for both $z=0$ and $z$ at the best-fit value.  For $z=0$ the model is excessive over the measured flux in one channel between $6.4-6.5$ keV, and only allowing $z$ to take positive values has the effect of slightly skewing the Fe emission in the model to more closely match the measured spectrum.      

The parameters of the power law are constant, with $\Gamma \sim 1.67$; consistent with the dominant power law found from  fitting the \emph{Chandra} data.  Finally, we obtained a best-fit with $z$, $\xi$ and the cross-calibration constant fixed, which we show with residuals in figure~\ref{fig:spectralfit} and present the corresponding parameters in table~\ref{tab:results}.  The model describes the SPI data well, with $\chi^2_\nu =46.17/48$.  Over the range $20-100$ keV, the flux of the power law component $F_{PL}=9.48_{-0.71}^{+0.71} \times 10^{-10}~{\rm erg~cm^{-2}~s^{-1}}$, while the reprocessed emission contributes a flux $F_{R}=3.15_{-0.68}^{+0.68} \times 10^{-10}~{\rm erg~cm^{-2}~s^{-1}}$, i.e. $\approx 25\%$ of the total emission. An important caveat to this is that there is degeneracy between $\xi $ and the normalisation $N_R$, and refitting with $\xi=55$ showed the flux fraction is consistently between $20-25\%$ of the total emission over $20-100$ keV.  

To translate this result to a form consistent with the literature, we fit the SPI spectra using the model {\bf pexrav} \citep{1995MNRAS.273..837M}, which describes the reflection of a seed cut-off power law spectra from a neutral accretion disc.  This has the key parameter $R$, which is an expression of the solid angle of the reflecting material illuminated by the central source ($0\rightarrow 2\pi$, normalised to 1).  We fix the $N_H$ and $\Gamma$ parameters to the dominant power law of the \emph{Chandra} spectra and also fix the inclination parameter to 0.82 \citep[corresponding to the cosine of $\sim 34^{circ}$, ][]{2007ApJ...671.1329N}.  We also varied $E_{cutoffpl}$ between values $500-1000$ keV (see \S~\ref{sec:cutrec}), finding that $R$ has negligible sensitivity to this parameter when $E_{cutoffpl}$ is above a few hundred keV.   We recover a value of $R=0.18^{+0.13}_{-0.11}$, which is consistent with previous attempts to measure the extent of reflection in Cen A with \emph{INTEGRAL} \citep[$R=0.12\pm0.1$,][]{2011A&A...531A..70B}.  

\begin{deluxetable}{lccccc}
\tabletypesize{\footnotesize}
\tablecaption{Best-fit parameters of {\bf phabs(powerlaw+reflionx)}}
\tablehead{
\colhead{$N_H$} &
\colhead{$\Gamma$} &
\colhead{$N_{PL}$} &
\colhead{$A_{Fe}$} &
\colhead{$N_{R}$} \\
\colhead{$\nh{22}$} &
\colhead{} &
\colhead{} &
\colhead{$\odot$} &
\colhead{$10^{-4}$} \\
}

\tablecolumns{6}
\startdata
$9.55^{-0.16}_{+0.16}$ & $1.67^{+0.02}_{-0.02}$ & $0.1139\pm0.0032$ & $0.69^{+0.24}_{-0.21}$ & $2.43^{+0.31}_{-0.30}$   \\
\enddata
\tablecomments{We fix $z$, $\xi$ and the multiplicative calibration constant $C$ (see text).  $N_{PL}$ and $N_{R}$ represent the normalisations of the power law and reflection components.} 
\label{tab:results}
\end{deluxetable}

\subsubsection{Testing for a high-energy cut-off in SPI data}

To test for the presence of a high-energy cut-off, we first defined a test statistic $\Delta\Gamma=\Gamma_{23-60}-\Gamma_{120-500}$; the difference in photon indices found from fitting an absorbed power law to the $23-60$ keV and $120-500$ keV ranges ($N_H$ always fixed at $\nh{23}$).  The motivation for this is to detect a change in spectral shape between low and high energy, as we would expect if there is a cut-off at a few hundred keV.  For the average Cen A spectrum, we calculate $\Delta\Gamma=-0.29$.   We then judge the unusualness of our test statistic based upon simulated spectra produced using the XSPEC command \emph{fakeit}.  

In the first instance, we simulate 1000 spectra based on the absorbed power law fit to the Cen A spectrum ($\Gamma=1.81$) and calculated $\Delta\Gamma$ for each of the spectra.  Only $3.2\%$ of spectra had a more negative $\Delta\Gamma$ than observed.  The $1\sigma$ Poisson uncertainty associated with a bin containing 32 counts is $\sim5.6$, and so we can confidently state that the Cen A spectrum lies outside $95\%$ of the distribution.  This is statistical evidence that the power law alone is not the correct description of the spectral shape.

The next stage was to simulate thousands of cut-off power law spectra with a range of $E_{cutoff}$ covering several plausible values for thermal Compton spectra; $200-500$ keV in steps of 100 keV.  We found that $\Delta\Gamma$ is sensitive to our choice of $E_{cutoff}$.  For $E_{cutoff}=200$ keV, we find that only $3.7\%$ of simulated spectra have $\Delta\Gamma$ as negative as observed, and this increases to $7\%$,$16\%$, $25\%$ for $E_{cutoff}=300$ to $500$ keV.  This suggests that a change in the spectral shape at high energy is a reasonable explanation for our observed $\Delta \Gamma$.  

Finally, we checked the sensitivity of $\Delta \Gamma$ to the presence of reflection in the spectrum, by simulating spectra as above, but with a {\bf reflionx} component contributing $30\%$ of the 23-60 keV flux.  Our testing showed that distribution of  $\Delta \Gamma$ varied very little with increased reflection flux, and even at $30\%$ contribution could only produce a result as the measured $\Delta \Gamma$ in fewer than $9\%$ of simulated spectra.

\subsubsection{Recoverability of $E_{cutoff}$}
\label{sec:cutrec}
To further investigate the location of a high energy rollover in the spectrum, we must first understand our ability to retrieve the true cut-off energy from fitting simplified spectral models.  This is of particular interest because all AGN observed with \emph{INTEGRAL} have lower quality data than Cen A, and fitting simple $3-4$ parameter models is the standard method of phenomenologically parametrising the data \citep[some or all sources in e.g.][]{2005A&A...444..431S,2008MNRAS.389.1360M,2009A&A...505..417B}.

\begin{figure}
\begin{center}
\includegraphics[width=0.8\hsize]{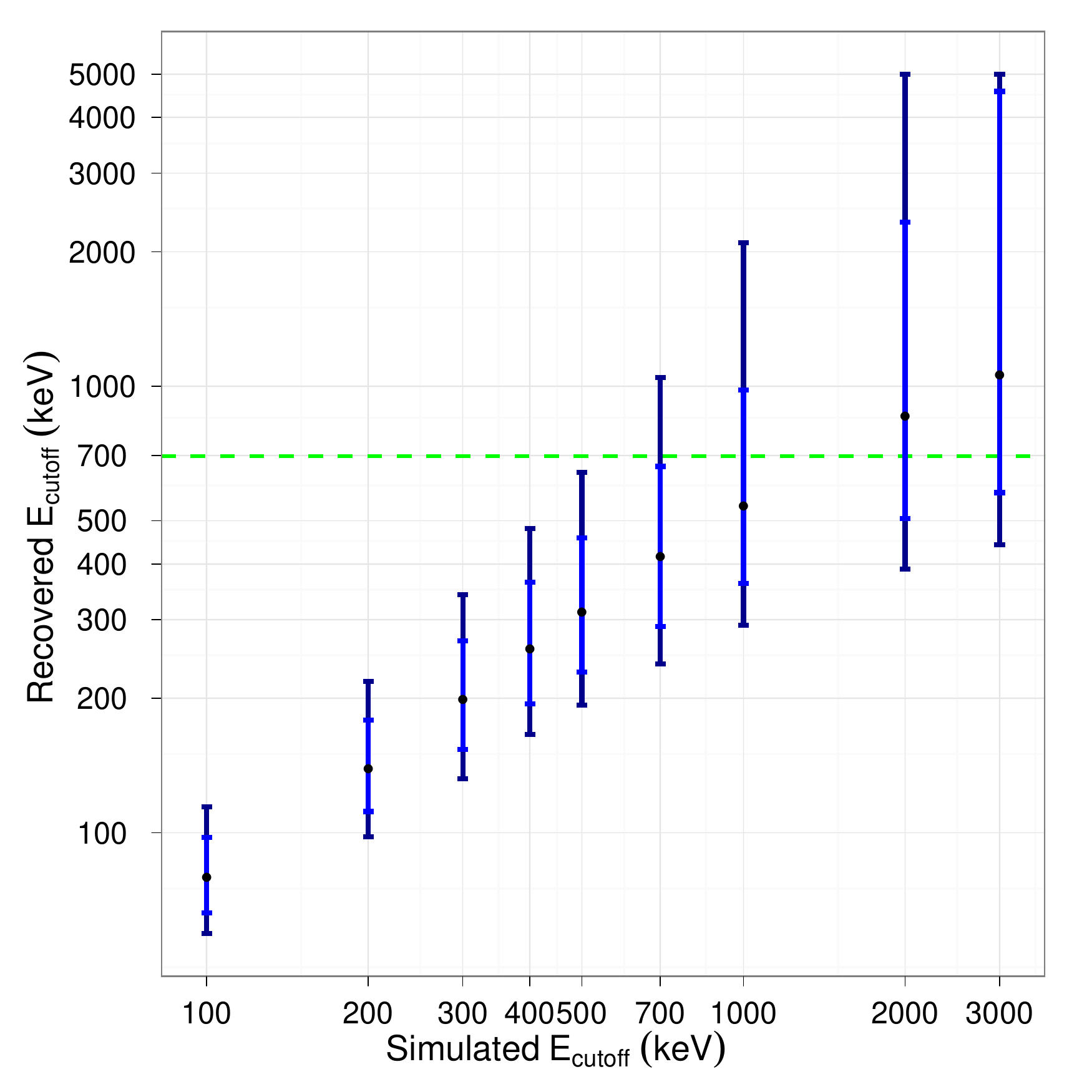} 
\caption{The $90\%$ ranges of recovered $E_{cutoff}$ found from fitting an absorbed cut-off power law (\emph{phabs(cutoffpl)}) to $10^4$ simulated spectra.  We also indicate the inner $68\%$ confidence to highlight the skewed natured of the distribution.  The simulated spectra possess a weak reflection component similar to that found in the Cen A average spectrum (such that the combined simulated model is \emph{phabs[cutoffpl+reflionx]}).  The recovered values are systematically lower than the true values used in the simulations.  The green dashed line indicates the measured $E_{cutoff}$ found from fitting the average spectrum of Cen A with an absorbed cut-off power law.  \label{fig:cutofflim}}
\end{center}
\end{figure}

For a range of $E_{cutoff}$ we simulated $10^4$ spectra based on our best-fit for Cen A (table~\ref{tab:results}), but with a cut-off present in the spectrum and with reflection contributing $\approx 10\%$ of the $23-60$ keV flux. We then fit the simulated spectra with an absorbed cut-off power law model ({\bf phabs $\times$ cutoffpl}).  In figure~\ref{fig:cutofflim} we present $68\%$ and $90\%$ confidence bars based on the distribution of recovered $E_{cutoff}$ against the actual values.  We present the two confidence bars to illustrate the skewed nature of the distributions.  In \emph{XSPEC}, the maximum possible value of $E_{cutoff}$ allowed was set to 5 MeV, and the fits would occasionally `peg' at this value for the higher values of $E_{cutoff}$.  It is clear that the fits systematically underestimate $E_{cutoff}$ when there is reflection present in the spectrum but ignored in the spectral fitting process. We indicate the recovered $E_{cutoff}$ from fitting the average Cen A spectrum with an absorbed cut-off power law (dashed horizontal line), $E_{cutoff}\sim 700$ keV.  This measurement lies outside of the main $90\%$ of the $E{cutoff}$ distribution when the actual $E_{cutoff} \leq 500$ keV.

\begin{figure}
\begin{center}
\includegraphics[width=0.8\hsize]{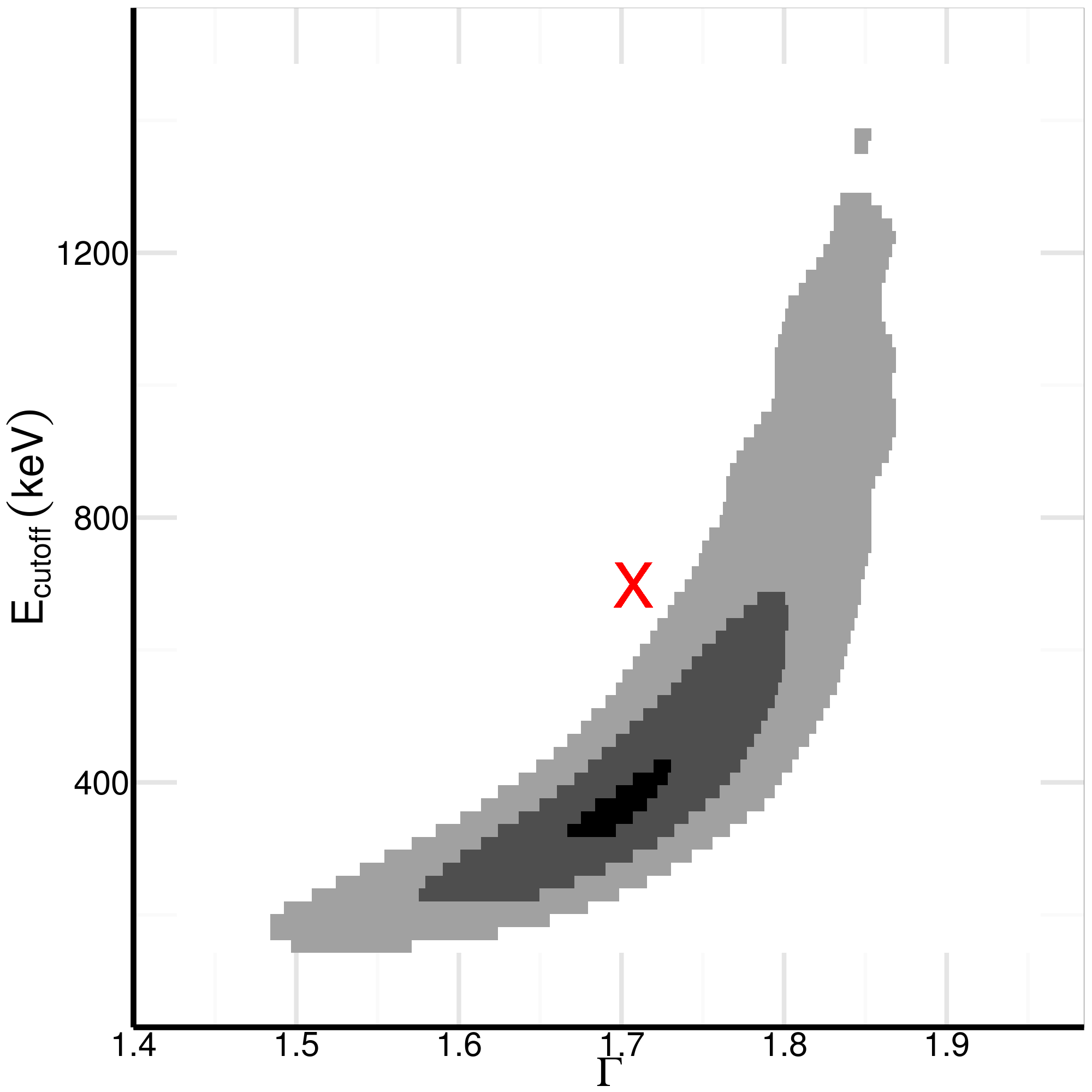}
\caption{The $95\%$, $68\%$ and $10\%$ confidence regions in the $E_{cutoff}-\Gamma$ plane found from fitting simulated spectra that possess a 700 keV cut-off.  Note that for the best-fits from $\sim 0.05\%$ of simulations $E_{cutoff}$ `pegged'  at the maximal value permitted, 5000 keV. For clarity, these data are not shown, but were factored into the calculation of each confidence region.  The red cross denotes the \emph{phabs(cutoffpl)} best-fit value for the average spectrum from Cen A.\label{fig:cutofflim2}}
\end{center}
\end{figure}

To estimate a lower-limit to the energy of a possible cut-off in the Cen A spectrum, we compare the measured $\Gamma-E_{cutoff}$ from fitting with an absorbed cut-off power law to the spectra simulated with $E_{cutoff}=700$ keV.  By computing the $95\%$ confidence region in the $\Gamma-E_{cutoff}$ plane (figure~\ref{fig:cutofflim2}), we find that the measured $\Gamma,E_{cutoff}$ excludes a cut-off energy of 700 keV with $>95\%$ confidence.




\section{Discussion}
We analysed the average hard X-ray to soft $\gamma$-ray spectrum of Cen A observed with {INTEGRAL} SPI, with an aim to saying more about the contribution of reprocessed emission to the spectrum and the existence of an high-energy cut-off.  These features have important implications for the geometry and accretion physics of the source.   This study makes use of 10 years of observations, and we present the lightcurve of this period in figure~\ref{fig:lc}.  The most striking feature of the lightcurve is the systematic increase in intensity between MJD 54000 and MJD 55000. Contemporaneous RXTE observations also recorded such an increase \citep{2011ApJ...733...23R}.  Based on the SPI lightcurve, the hard X-ray emission from the Cen A nucleus can vary in intensity by as much as a factor of four on timescales of years, and by a factor of two over a period of a few months, while the spectral shape remains constant, consistent with the RXTE results.


We revisit the earlier Chandra grating spectra from Cen A.  This was motivated by the need to model the Fe $K\alpha$ complex consistently with the energy range of the Compton hump.    We examine the consistency between these low and high energy datasets by fitting the normalisations of the dominant power law reported by \cite{2004ApJ...612..786E}, and show that this inadequately models the high-energy spectrum (figure~\ref{fig:spectralfit}), the shape of the residuals indicating an excess between $20-50$ keV.  We propose that this excess is physical evidence of the so-called Compton hump, resulting from a reprocessing of the central X-ray emission.  To model the excess at $20-50$ keV and 6-7 keV in a consistent way as reflection from a neutral, distant absorber we employed the model \emph{reflionx}, first to demonstrate the low ionisation state of the reflecting medium, as expected for neutral material before fixing the ionisation parameter to the minimal value in subsequent fitting.  In this way, we obtain a good description of the low- and high-energy spectra (figure~\ref{fig:spectralfit}). The best-fit model parameters (table~\ref{tab:results}) suggest that reflection may account for as much as $20-25\%$ of the $20-100$ keV flux, and that the reflecting may have sub-solar abundances ($A_{Fe}=0.69^{+0.21}_{-0.24}$).  The photon index dominant power law component is consistent with $\Gamma=1.67$, as was found by \cite{2004ApJ...612..786E}, and so fitting the reflection component directly describes the excess in spectral shape demonstrated in figure~\ref{fig:powerlawres}.

It is well-established that the spectral shape of Cen A changes between the X-ray and hard $\gamma$-ray regime \citep{2010ApJ...719.1433A}, with some authors proposing that this takes the form of a cut-off defined by the characteristic temperature of a thermal Comptonizing plasma \citep[e.g.][]{2011A&A...531A..70B} or more gradual spectral curvature as would be expected for SSC emission, as evidenced from the broader SED \citep[][]{2001MNRAS.324L..33C,2010ApJ...719.1433A}. The lack of ionised or broadened emission lines indicates that there is not a significant emission component arising from the disk \citep{2000ApJ...528..276M,2004ApJ...612..786E}, which is in agreement with our conclusion that the jet is the principal emitter.

The $\Delta \Gamma$ statistic, which we defined as the difference in spectral slope between low ($23-60$ keV) and high ($120-600$ keV) energy, is very sensitive to the presence of a cut-off.  Based on $\Delta \Gamma$ we show that a simple power law is an unlikely description of the spectrum at high-energy (despite an acceptable $\chi^2_\nu$, \S~\ref{sec:specfit}), and the measured value of $\Delta \Gamma$ can be reproduced from a cut-off at a few hundred keV. Absorbed cut-off power law fits to the average spectrum recovered $E_{cutoff}\sim700$ keV.  By simulating spectra dominated by a $\Gamma~\sim 1.67$ cut-off power law with a comparatively small reflection component,  over a range of $E_{cutoff}$ and then fitting a simple cut-off power law recovers parameters that lie outside of $95\%$ confidence region for $E_{cutoff}=700$ (figure~\ref{fig:cutofflim2}), all the more so for smaller $E_{cutoff}$ (figures~\ref{fig:cutofflim}).  Recalling that for a thermal Comptonizing plasma $E_{cutoff} \sim 2-3 \times kT_e$ \citep{1980A&A....86..121S}, this suggests $kT_e > 230$ keV, and even a conservative $E_{cutoff}\sim500$ keV leads to $kT_e\sim 160-250$ keV.  \cite{2013MNRAS.433.1687M} suggest a reasonable upper-limit of $kT_e < 65 - 100$ keV for average thermal Compton emission from all AGN, based on cosmic diffuse background measurements \citep{2007A&A...463...79G}, and we suggest that this argues against this being the primary source of hard X-rays in Cen A.  Our results support the picture of Cen A where the broad SED can be described by SSC emission \citep{2001MNRAS.324L..33C,2010ApJ...719.1433A}, consistent with other FR I galaxies  that are considered to be BL Lac objects observed along a line-of-sight that is not aligned to the jet axis \citep{2009ApJ...699...31A,2009ApJ...707...55A}.  

Finally, our results from fitting a solitary absorbed cut-off power law to spectra that are dominated by a cut-off power law but also  contain a modest amount of reflection shows that $E_{cutoff}$ is systematically underestimated.  This effect increases with increased reflection.  As it is often the case that studying AGN with \emph{INTEGRAL} requires fitting an absorbed cut-off power law to the data to try to obtain a general picture of the thermal plasma in AGN, we believe that the inferred $kT_e$ will have been over-estimated for many sources where reflection has not been taken into account.

We thank the anonymous referee for their careful reading of the manuscript and for many helpful comments that greatly improved the readability of this paper.

The INTEGRAL SPI project is supported under the responsibility and leadership of CNES. We are grateful to
ASI, CEA, CNES, DLR, ESA, INTA, NASA, and OSTC for support.

\bibliography{cena}{}
\bibliographystyle{apj}

\end{document}